\documentclass{article}

\usepackage{amsmath}
\usepackage{amssymb}
\usepackage{amsthm}
\usepackage{mytheorems}
\usepackage{latexsym} 
\usepackage[all]{xypic}
\usepackage{textlogic}

\input macros.tex

\begin{document}

\title{%
  A Nice Labelling for \\
  Tree-Like Event Structures of Degree $3$\thanks{%
    Research supported
    by the \emph{Agence Nationale de la Recherche}, project SOAPDC
    no. JC05-57373} \\
  (Extended Version)
}

\author{Luigi Santocanale \\
  LIF, Universit\'e Aix-Marseille,
  \\
  39 rue F. Joliot Curie, 13453 Marseille Cedex 13, France \\
  \texttt{luigi.santocanale@lif.univ-mrs.fr}
}%

\maketitle

\begin{abstract}
We address the problem of finding nice labellings for event structures
of degree $3$. We develop a minimum theory by which we prove that the
index of an event structure of degree $3$ is bounded by a linear
function of the height.  The main theorem of the paper states that
event structures of degree $3$ whose causality order is a tree have a
nice labelling with $3$ colors.  We exemplify how to use this theorem
to construct upper bounds for the index of other event
structures of degree $3$.


\end{abstract}


\section{Introduction}


Event structures, introduced in \cite{eventstructures1}, are nowadays
a widely recognized model of true concurrent computation and have
found many uses since then. They are an intermediate abstract model
that makes it possible to relate  more concrete models such as
Petri Nets or higher dimensional 
automata \cite{winskelnielsen}. They
provide formal semantics of process calculi \cite{winskelccs,varacca}.
More recently, logicians became interested in event structures with the
aim of constructing models of proof systems that are invariant under
the equalities induced by the cut elimination procedure
\cite{faggian,mellies}.


Our interest for event structures stems from the fact that they
combine distinct approaches to the modeling of concurrent computation.
On one side, language theorists have developed the theory of partially
commutative monoids \cite{bookoftraces} as the basic language to
approach concurrency.  Classes of automata that properly model
concurrent processes -- such as asynchronous automata \cite{zielonka}
or concurrent automata \cite{droste} -- have been studied as part of
this theory.  On the other hand, the framework of domain theory and,
ultimately, order theoretic ideas have often been proposed as the
proper tools to handle concurrency, see for example \cite{pratt}.  In
this paper we pursue a combinatorial problem that lies at the
intersection of these two approaches.  It is the problem of finding
nice labellings for event structures of fixed degree. To our
knowledge, this problem has not been investigated any longer since it
was posed in \cite{rozoy1,rozoy2} and partially solved in
\cite{rozoy}.


Let us recall that an event structure is made up of a set of local
events $E$ partially ordered by the causality relation $\leq$.
Causally independent events may also be in the conflict relation
$\confl$. A global state of the computation, comprehensive of its
history, is modeled as a subset of events, lower closed w.r.t. the
causality relation, which also is an independent set w.r.t. the
conflict relation.  These global states may be organized into a poset,
the domain of an event structure, representing all the concurrent
non-deterministic executions. The Hasse diagram of this poset codes
the state-transition graph of the event structure as an abstract
process.  By labeling the transitions of this graph with letters from
some alphabet, we can enrich the graph with the structure of a
deterministic concurrent automaton.  The nice labelling problem asks
to find a labelling that uses an alphabet of minimum size. The
size of this alphabet is called the \emph{index} of the event
structure.

The problem is
actually equivalent to a graph coloring problem in that we can
associate to an event structure a graph, of which we are asked to
compute the chromatic number.
The \emph{degree} of an event structure is the maximum out-degree of a
node in the Hasse diagram of the associated domain, that is, the
maximum number of upper covers of some element. Under the graph
theoretic translation of the problem, the degree coincides with the
clique number, and therefore it is a lower bound for the cardinality
of a solution.  A main contribution in \cite{rozoy} was to prove that
event structures of degree $2$ have index $2$, i.e. they posses a nice
labelling with $2$ letters.
On the other hand, it was proved there that event structures of higher
degrees may require strictly more letters than the degree.

The labelling problem may be thought to be a generalization of the
problem of covering a poset by disjoint chains.  Dilworth's Theorem
\cite{dilworth} states that the minimal cardinality of such a cover
equals the maximal cardinality of an antichain. This theorem and the
results of \cite{rozoy} constitute the few knowledge on the problem
presently available to us. For example, we cannot state that there is
some fixed $k > n$ for which every event structure of degree $n$ has a
nice labelling with at most $k$ letters.  In light of standard graph
theoretic results \cite{mycielski}, the above statement should not be
taken for granted.


We present here some first results on the nice labelling problem for
event structures of degree $3$.  We develop a minimum theory that
shows that the graph of a degree $3$ event structure, when restricted
to an antichain, is almost acyclic and can be colored with $3$
letters.  This observation allows to construct an upper bound to the
labelling number of such event structure as a linear function of its
height.  We focus then on event structures whose causality order is a
tree or a forest. Intuitively, these tree-like event structures
represent concurrent systems where processes are only allowed to fork
or to take local nondeterministic choices.  Our main theorem states
that tree-like event structures of degree $3$ have a nice labelling
with $3$ letters.
Finally, we suggest how to use this and other theorems to construct
upper bounds for the index of other event structures of degree
$3$. These general upper bounds depend on some parameter. To exemplify
the scope of theory, we prove a constant upper bound on a simple class
of degree $3$ event structures.

The general question we address, whether there exists a finite common
upper bound to the indexes of event structures of degree $3$, remains
open. Conscious that this question might be difficult to answer in its
full generality -- as  usual for graph coloring problems -- we
felt worth to present these partial results and to encourage other
researchers to pursue this and other combinatorial problems that arise
from concurrency.
Let us mention why these problems deserve an in-depth
investigation. The theory of event structures is presently being
applied to automated verification of systems.  Some model checkers --
see for example \cite{kit} and \cite{poem} -- make explicit use of
trace theory and of the theory of partially ordered sets to represent
the state space of a concurrent system. The combinatorics of posets is
then exploited to achieve an  efficient exploration of the global
states of concurrent systems
\cite{mcmillan,partialorderreduction,esparza}.  Thus, having a solid
theoretical understanding of such combinatorics is a prerequisite and
a complement for designing efficient algorithms for these kind of
tools.


The paper is structured as follows. After recalling the order
theoretic concepts we shall use, we introduce event structures and the
nice labelling problem  in section \ref{sec:theproblem}.
In section \ref{sec:antichains} we develop the first properties of
event structures of degree $3$. As a result, we devise an upper bound
for the labelling number of such event structures as a linear function
of the height.
In section \ref{sec:thetheorem} we present our main result stating
that event structures whose underlying order is a tree may be
labeled with $3$ colors.
In section \ref{sec:w2} we develop a general approach to construct
upper bounds to the labelling number of event structures of degree
$3$.  Using this approach and the results of the previous section, we
compute a constant upper bound for a class of degree $3$ event
structures that have some simplifying properties, that consequently we
call simple.

\noindent%
\emph{Acknowledgement.} %
We would like to thank R\'emi Morin for introducing and guiding us to
and through the theory of concurrency.
 

\subsubsection*{Order Theoretic Preliminaries.}
\label{sec:notation}

Let us anticipate that part of an event structure is a set $E$ of
events which is partially ordered by the causality relation $\leq$. As
in this paper we shall heavily rely on order theoretic concepts, we
introduce them here together with the notation we shall use.  All
these concepts will apply to the poset $\langle E,\leq\rangle$ of an
event structure.
 
A finite poset is a pair $\langle P,\leq \rangle$ where $P$ is a
finite set and $\leq$ is a reflexive, transitive and antisymmetric
relation on $P$.  A subset $X \subseteq P$ is a \emph{lower set} if $y
\leq x \in X$ implies $y \in X$. If $Y \subseteq P$, then we denote by
$\idealof{Y}$ the least lower set containing $Y$. Explicitly,
$\idealof{Y} =\set{x \in P\mid \exists y \in Y \tst x \leq y}$.
Two elements $x,y \in P$ are \emph{comparable} if and only if either
$x \leq y$ or $y \leq x$. We write $x \comp y$ to mean that $x,y$ are
comparable.  A \emph{chain} is sequence $x_{0},\ldots ,x_{n}$ of
elements of $P$ such that $x_{0} < x_{1} < \ldots < x_{n}$. The
integer $n$ is the length of the chain.  
The \emph{height} of an
element $x \in P$, noted $\height(x)$, is the length of the longest
chain in $\ideal{x}$. The height of $P$ is $\max \set{\height(x)\mid 
  x \in P}$.  
Let us write $x < y$ if $x \leq y$ but $x \neq y$.  An
\emph{antichain} is a subset $X \subseteq P$ such that $x \not\comp y$
for each pair of distinct $x,y \in X$.  The \emph{width} of $\langle
P, \leq \rangle$, noted $\width(P,\leq)$, is the integer $\max
\set{\card(A)\mid  A \text{ is an antichain}}$.
If the interval $\set{z \in P\mid  x \leq z \leq y}$ is the two
elements set $\set{x,y}$, then we say that $x$ is a \emph{lower cover}
of $y$ or that $y$ is \emph{an upper cover} of $x$. We denote this
relation by $x \lcover y$. The Hasse diagram of $\langle P,\leq
\rangle$ is the directed graph $\langle P,\lcover \rangle$.  For $x
\in P$, the \emph{degree} of $x$, noted $\cl(x)$, is the number of
upper covers of $x$. That is, the degree of $x$ is the outdegree of
$x$ in the Hasse diagram. The degree of $\langle P, \leq \rangle$,
noted $\cl(P,\leq)$, is the integer $\max \set{\cl(x)\mid x \in P}$.
We shall denote by $\fat(x)$ the number of lower covers of $x$ (i.e.
the indegree of $x$ in the Hasse diagram).  The poset $\langle P,\leq
\rangle$ is \emph{graded} if $x \lcover y$ implies $\height(y) =
\height(x) + 1$.

 
\section{Event Structures and the Nice Labelling Problem}
\label{sec:theproblem}

Event structures are a basic model of concurrency introduced in
\cite{eventstructures1}. The definition we present here is from
\cite{winskelnielsen}.
\begin{definition}
  An \emph{event structure} is a triple $\E = \langle
  E,\leq,\CNF\rangle$ such that
  \begin{enumerate}
  \item $\langle E,\leq\rangle$ is a poset, such that for each $x \in
    E$ the lower set $\ideal{x}$ is finite,
  \item $\CNF$ is a collection of subsets of $E$ such that:
    \begin{itemize}
    \item $\set{x} \in \CNF$ for each $x \in E$,
    \item $X \subseteq Y \in \CNF$ implies $X \in \CNF$,
    \item $X \in
      \CNF$ implies $\idealof{X} \in \CNF$.
    \end{itemize}
  \end{enumerate}
\end{definition}
In  this paper we shall consider finite event structures only, so that
that $\ideal{x}$ is always finite.
The order $\leq$ of an event structure $\E$ is known as the
\emph{causality} relation between events. The collection $\CNF$ is
known as the set of configurations of $\E$. A configuration $X \in
\CNF$ of causally unrelated events -- that is, an antichain w.r.t.
$\leq$ -- is a sort of snapshot of the global state of some
distributed computation. A snapshot $X$ may be transformed into a
description of the computation that takes into account its history.
This is done by adding to $X$ the events that causally have determined
events in $X$. That is, the history-aware description is the lower set
$\idealof{X}$ generated by $X$. We shall be particularly interested in
the collection of history-aware configurations, defined as
\begin{align*}
  \Hi & = \set{Y \in \CNF \mid \,\idealof Y = Y}\,.
\end{align*}
Observe that $X \in \CNF$ if and only if $\idealof X \in \Hi$, so that
we do not loose information if we focus on history-aware
configurations.

Two events $x,y \in E$ are said to be \emph{concurrent} if $x \uncomp
y$ and there exists $X \in \CNF$ such that $x,y \in X$.  We shall
write $x \conc y$ to mean that $x,y$ are concurrent.  It is useful to
introduce a weakened version of the concurrency relation where we
allow events to be comparable: $x \wconc y$ if and only if $x \conc y$
or $x \comp y$. Equivalently, $x \wconc y$ if and only if there exists
$X \in \CNF$ such that $x,y \in X$.
The set of configurations that arise from many concrete models is
completely determined by the concurrency relation. For example, this
is the case for event structures that code the behavior of $1$-safe
Petri-nets.
\begin{definition}
  An event structure $\E$ is \emph{coherent} if $\CNF$ is the set
  of cliques of the weak concurrency relation: $X \in \CNF$ if
  and only if $x \wconc y$ for every pair of events $x,y \in X$.
\end{definition}
Coherent event structures are also known as \emph{event structures
  with binary conflict}. To understand this name, let us explicitely
introduce the conflict relation and two other derived relations:
\begin{enumerate}
\item \emph{Conflict}: $x \confl y$ if and only if $x \ncomp y$ and $x
  \nconc y$.
\item \emph{Minimal conflict}: $x \mconfl y$ if and only (i) $x \confl
  y$, (ii) $x' < x$ implies $x' \wconc y$, and (iii) $y' < y$ implies
  $x \wconc y'$.
\item \emph{Orthogonality}: $x \orth y$ if and only if $x \mconfl y$
  or $x \conc y$.
\end{enumerate}
A coherent event structure is completely described by the triple
$\langle E,\leq,\confl\rangle$ where the conflict relation is
symmetric and irreflexive, and moreover is such that $x \confl z$
whenever $x \confl y$ and $y\leq z$.

The concurrency relation, being the restriction to uncomparable
elements of the complement of the conflict relation, satisfies the
following conditions:
\begin{enumerate}
\item 
  $x \conc y$ implies $x \ncomp y$,
\item 
  $x \conc y$ and $z
  \leq x$ implies $z \conc y$ or $z \leq y$.
\end{enumerate}
In this paper we deal mainly with coherent event structures  and,
unless explicitly stated, event structure will be a synonym for
coherent event structure.

\subsubsection*{Coloring the Graph of an Event Structure}
The orthogonality relation clearly is symmetric. Thus, by identifying
an ordered pair $(x,y)$ such that $x \orth y$ with the unordered pair
$\couple{x,y}$, we shall focus on the undirected graph induced by the
orthogonality\footnote{%
  Let us observe that two orthogonal events are called independent
  in \cite{rozoy}. An independent set in the complement undirected graph
  $\langle V, E^{c} \rangle$ is a clique of the graph $\langle V, E
  \rangle$, thus explaining terminology used in \cite{rozoy}.  In this
  paper we shall focus on the structural properties of the graph
  $\G(\E) = \langle E, \orth \rangle$ and not of its complement, and
  therefore we prefer to deviate from the existing terminology.  }
relation.
This graph, formally defined by
\begin{align*}
  \G(\E) & = \langle E, \orth \rangle \,,
\end{align*}
will be called the graph of $\E$.
Let us  list some properties of the orthogonality relation:
\begin{enumerate}
\item $x \orth y$ if and only if (i) $x \not\comp y$, (ii) $x ' < x$
  implies $x' \wconc y$, (iii) $y' < y$ implies $x \wconc y'$,
\item if $x \orth y$ and $z \leq x$ then $z \orth y$
  or $z \leq y$.
\end{enumerate}
Together with the following property:
\begin{enumerate}
  \setcounter{enumi}{2}
\item if $x \confl y$ then there exists $x' \leq x$ and $y' \leq y$
  such that $x' \mconfl y'$,
\end{enumerate}
which ties up the conflict relation with the orthogonality through the
minimal conflict, these properties shall be our main working tool. We
leave the proof of them as an exercise for the reader.


\begin{definition}
  A \emph{nice labelling} of an event structure $\E$ is a coloring of
  the graph $\G(\E)$. That is, it is a pair $(\lambda,\Sigma)$ with
  $\Sigma$ is a finite alphabet and $\lambda : E \rTo \Sigma$ such
  that $\lambda(x) \neq \lambda(y)$ whenever $x \orth y$.
\end{definition}

For a graph $G$, let $\chr(G)$ denote its chromatic number and let
$\cl(G)$ be its clique number, i.e. the size of the largest clique of
$G$.  
\begin{definition}
  The \emph{degree} of $\E$, $\cl(\E)$, is the clique number of
  $\G(\E)$, i.e. the number $\cl(\G(\E))$.  The \emph{index} of $\E$,
  $\chr(\E)$, is the chromatic number of $\G(\E)$, i.e. the number
  $\chr(\G(\E))$.
\end{definition}
\breath %
The \emph{\textbf{nice labelling problem}} asks to compute $\chr(\E)$
for a given event structure $\E$.  It was shown to be an NP-complete
problem in \cite{rozoy}. The graph theoretic definition of the index
and of the degree makes it clear that $\cl(\E) \leq \chr(\E)$.  More
generally, given a class ${\cal K}$ of event structures, the
\emph{\textbf{nice labelling problem for the class ${\cal K}$}} asks
to compute the index of ${\cal K}$, defined as
\begin{align*}
  \chr({\cal K}) & = \max \set{\chr(\E) \mid \E \in {\cal K}}\,.
\end{align*}
Of course, $\chr({\cal K})$ might not be a finite number.  A necessary
condition for the relation $\chr({\cal K}) < \infty$ to hold is the
existence of a finite upper bound on the size of cliques of the graphs
$\G(\E)$ with $\E \in {\cal K}$.  Thus, of particular interest are the
classes of event structures ${\cal K}_{n}$ defined by
\begin{align*}
  {\cal K}_{n} & = \set{\E \mid \cl(\E)
    \leq n }\,.
\end{align*}
It is time to recall the known results on the nice labelling problem
for classes of event structures. The first one is the celebrated
Dilworth's theorem. 
\begin{theorem}[Dilworth \cite{dilworth}]
  If the conflict relation of $\E$ is empty, then $\chr(\E) =\cl(\E)$.
\end{theorem}
As a matter of fact, if the conflict relation is empty, then $x \orth
y$ if and only if $x,y$ are not comparable, so that nice labellings of
$\E$ are in bijection with coverings of the poset $\langle E,
\leq\rangle$ by disjoint chains.
Notice next that the conflict relation of $\E$ is empty if and only if
there is no pair of events $x,y \in E$ such that $x \mconfl y$,
i.e. that are in minimal conflict.  Dilworth's Theorem, as a statement
about event structures with a limited number of minimal conflicts, has
the following generalization:
\begin{theorem}[Assous et al. \cite{rozoy}]
  If
  \begin{align*}
    {\cal K}_{n,m} & = \set{\E \in {\cal K}_{n} \mid 
      \card \set{(x,y)\mid x \mconfl y} \leq m}\,,
  \end{align*}
  then $\chr({\cal K}_{n,m})  < \infty$.
\end{theorem}
Dilworth's theorem, as a particular case of the previous theorem,
states that ${\cal K}_{n,0} = n$.  The next result, dealing with event
structures of degree $2$, has been our motivating staring point.
\begin{theorem}[Assous et al. \cite{rozoy}]
  $\chr({\cal K}_{2}) = 2$ and $\cl({\cal K}_{n}) > n$ for $n > 2$.
\end{theorem}

\subsubsection*{Computational Interpretation of the Nice Labelling
  Problem}
The rest of this section is meant to clarify the role of the
orthogonality relation and of the graph $\G(\E)$. The computational
interpretation we shall give is part of the folklore in concurrency
theory, see for example \cite{darondeau}, but it is worth recalling.
Let us first review the definition of the domain of an event
structure.
\begin{definition}
  The domain $\D(\E)$ of an event structure $\E = \langle E,\leq,\CNF
  \rangle$ is the poset $\langle \Hi, \subseteq \rangle$, where $\Hi$
  is the collection of history-aware configurations of $\E$.
\end{definition}
Following a standard axiomatization in theoretical computer science
$\D(\E)$ is a \emph{stable $L$-domain}, see
\cite{winskelnielsen,droste:domains}.  This property roughly means
that $\D(\E)$ almost is a \emph{distributive} lattice. Let us stress
this point, as most of the following considerations are elementary
observations of the theory of distributive lattices.

The collection $\Hi$ being closed under binary intersections, the
poset $\D(\E)$ is a finite meet semilattice -- or a chopped lattice as
defined in \cite[Chapter 4]{gratzer}. It is distributive in the
following sense:\footnote{%
  Usually, a meet semilattice is said to be distributive if its filter
  completion is a distributive lattice, see for example
  \cite{wehrung}.  } the equation $z \land (x \vee y) = (z \land x)
\vee (z \vee y)$ is satisfied whenever $x \vee y$, the least upper
bound of $\set{x,y}$, exists. The following Lemma asserts that finite
distributive meet semilattices are essentially the same structures as
the domains of (possibly not coherent) event structures.
\begin{lemma}
  Every finite distributive meet semilattice is isomorphic to the
  domain of an event structure.
\end{lemma}
\begin{proof}
  Since the ideas on which the proof relies are well known, we only
  sketch it.
  Let $L$ be a finite distributive meet semilattice, say that $x \in
  L$ is prime if it has a unique lower cover and denote by $J(L)$ the
  set of prime elements of $L$.  As usual from lattice theory, argue
  that $x \leq z \vee y$ implies $x \leq z$ or $x \leq y$ whenever $x
  \in J(L)$ and the least upper bound $z \vee y$ exists.  For $X
  \subseteq J(L)$ say that $X \in \CNF$ if the least upper bound of
  $X$ exists in $L$. Let then $\E = \langle J(L),\leq,\CNF \rangle$,
  it is a standard exercise to prove that $\D(\E)$ is order isomorphic
  to $L$.
\end{proof}
A lower set in $\Hi$ represents a state of the global computation,
comprehensive of its history.  
For
$I,J \in \Hi$, $I \subseteq J$ intuitively means that the global
state $J$ may take place after the global state $I$.  The Hasse
diagram of $\D(\E)$ therefore represents the state-transition graph of
$\E$ as a process. 
We obtain a representation of
the process $\E$ as an automaton if we color the edges of the Hasse
diagram by letters of some alphabet.  It is quite natural, however, to
ask this coloring to satisfy the following conditions.
\begin{description}
\item[Determinism:] transitions outgoing from the same state have
  different colors.
\item[Concurrency:] every square of the diagram has to be colored
  according to the following pattern, suggesting that actions
  $\sigma,\tau$ may take place in parallel:
  \begin{align}
    &\xygraph{%
      []*+{I}="a" ( [ru]*+{J_{1}}="b", [lu]*+{J_{0}}="c" )
      "a"[u(2)]*+{J_{0} \cup J_{1}}="d" "a"(-"b"^{\sigma},-"c"_{\tau}) "b"-"d"_{\tau}
      "c"-"d"^{\sigma} }
    & 
    \tilde{\lambda}(I \lcover J_{1})
  & = \sigma = \tilde{\lambda}(J_{0} \lcover J_{0} \cup J_{1})\,.
\label{eq:concurrency}
  \end{align}
\end{description}
Let us analyze what it means for an edge-coloring to be concurrent.
Consider that if $I\lcover J$ is an edge of the Hasse diagram of
$\D(\E)$, then $J = I \cup \set{x}$ for some $x \in E \setminus I$
such that $y \in E$ whenever $y < x$.  Thus, if we start from a
labelling $\lambda : E \rTo \Sigma$ and define $\tilde{\lambda}(I
\lcover I \cup \set{x}) = \lambda(x)$, then condition
\eqref{eq:concurrency} is fulfilled:
\begin{align*}
  \tilde{\lambda}(I \lcover J_{0}) & =
  \tilde{\lambda}(I \lcover I \cup \set{x_{0}}) = \lambda(x_{0}) \\
  & = \tilde{\lambda}(I \cup \set{x_{1}} \lcover  I \cup \set{x_{0}
    x_{1}}) 
   = \tilde{\lambda}(J_{1} \lcover  J_{0} \cup J_{1})\,.
\end{align*}
In order to see that every concurrent edge-coloring arise in this way,
observe that by translating down colors along opposite side of
concurrent squares as in \eqref{eq:concurrency}, a concurrent
edge-coloring is determined by the ideals in $\D(\E)$ with a unique
lower covers; these are of the form $\ideal{x}$. Thus we have
observed:
\begin{lemma}
  There is a bijection between concurrent edge-colorings of the Hasse
  diagram of $\D(\E)$ and functions $\lambda : E \rTo \Sigma$.
\end{lemma}

We analyze next how the condition on determism of a concurrent
edge-coloring transfers to a function $\lambda : E \rTo \Sigma$.
The following is the key Lemma to understand the role of the
orthogonality relation.
\begin{lemma}
  \label{lemma:degree}
  A set $\set{x_{1},\ldots ,x_{n}}$ is a clique of $\G(\E)$ if and
  only if there exists an history-aware configuration $I$
  susch that $I\lcover I \cup \set{x_{i}}$, $i = 1,\ldots ,n$, are
  distinct edges of the Hasse diagram of $\D(\E)$.
\end{lemma}
\begin{proof}
  Suppose that $I \cup \set{x_{i}}$ and $I \cup \set{x_{j}}$ are
  distinct upper covers of some $I$ in $\D(\E)$.  Then
  $\{x_{i},x_{j}\}$ is an antichain since $x_{i} \leq x_{j}$ implies
  that $I \cup \set{x_{i}} \subseteq I \cup \set{x_{j}}$.  If $x' <
  x_{i}$ then $x' \in I \subseteq I \cup \set{x_{j}}$. Since $I \cup
  \set{x_{j}}$ is a clique for the weak concurrency relation, then $x'
  \wconc x_{j}$.  Similarly $y' < x_{j}$ implies $x_{i} \wconc y'$ and
  therefore $x_{i} \orth x_{j}$. In particular, distinct upper covers
  of some $I$ give rise to a clique in $\G(\E)$.

  Conversely, let us suppose that $x_{i} \orthogonal x_{j}$ whenever
  $i \neq j$ and recall that $x_{i} \orthogonal x_{j}$ implies $x'
  \wconc y'$ for $x' < x$ and $y'< y$.  Thus, if we let $I =
  \bigcup_{i=1}^{n}\set{x'\mid x' < x_{i}}$, then $I \in \D(\E)$ and
  $I \cup \set{x_{i}} \in \D(\E)$ as well, for $i =1,\ldots ,n$. If $i
  \neq j$, then $x_{i},x_{j}$ are not comparable and therefore $I \cup
  \set{x_{i}}$ and $I \cup \set{x_{j}}$ are distinct upper covers of
  $I$.
\end{proof}
Let us remark that the Lemma strongly depends on $\E$ being a coherent
event structure.  The Lemma also implies that the degree of $\E$, that
is maximum size of a clique in $\G(\E)$, coincides with the the
maximum out-degree a configuration in the Hasse diagram of $\D(\E)$.
The degree of $\E$ is nothing else but the degree of the poset
$\D(\E)$ as defined on page \pageref{sec:notation}.  Considering the
case $n = 2$ in the statement of Lemma \ref{lemma:degree}, we deduce
the following Proposition:
\begin{proposition}
  There is a bijection between concurrent deterministic edge-colorings
  of the Hasse diagram of $\D(\E)$ and colorings the graph of
  $\G(\E)$.
\end{proposition}
Consequently, the size of a minimal alphabet by which we can transform
the Hasse diagram of $\D(\E)$ into a deterministic concurrent
automaton coincides with the chromatic  number of $\G(\E)$, what we
called the index of $\E$.

\section{Cycles and Antichains}
\label{sec:antichains}

From now on, in this and the following sections, $\E = \langle E,
\leq, \CNF\rangle$ will be a fixed coherent event structure of degree
at most $3$.  We begin our investigation of the nice labelling problem
for $\E$ by studying the restriction to an antichain of the graph
$\G(\E)$.
The main tool we shall use is the following Lemma, a straightforward
generalization of \cite[Lemma 2.2]{rozoy} to degree $3$. In
\cite{getco06} we proposed generalizations of this Lemma to higher
degrees and pointed out the geometrical flavor of the resulting
statements.
\begin{lemma}
  \label{lemma:avoided}
  Let $\set{x_{0},x_{1},x_{2}}, \set{x_{1},x_{2},x_{3}}$ be two size
  $3$ cliques in the graph $\G(\E)$ sharing the same face
  $\set{x_{1},x_{2}}$. Then $x_{0},x_{3}$ are comparable.
\end{lemma}
\begin{proof}
  Let us suppose that $x_{0},x_{3}$ are not comparable. It is not
  possible that $x_{0} \orth x_{3}$, since then we have a size $4$
  clique in the graph $\G(\E)$. Thus $x_{0} \confl x_{3}$ and we can
  find  $x'_{0}
  \leq x_{0}$ and $x'_{3} \leq x_{3}$ such that $x'_{0} \mconfl
  x'_{3}$.  We claim that $\set{x'_{0},x_{1},x_{2},x'_{3}}$ is a size
  $4$ clique in $\G(\E)$, thus reaching a contradiction.
  If $x'_{0} \north x_{1}$, then $x'_{0} \leq x_{1}$. However,
  $x'_{0} \leq x_{1} \orth x_{3}$ implies $x'_{0}\wconc x_{3}$, and
  henceforth $x'_{0} \wconc x'_{3}$. The latter relation contradicts
  $x'_{0} \mconfl x'_{3}$. Similalry, $x'_{0} \orth x_{2}$, $x'_{3}
  \orth x_{1}$, $x'_{3}\orth x_{2}$.
\end{proof}

We are going to improve the previous Lemma. To this goal, let us say
that a sequence $x_{0}x_{1}\ldots x_{n-1}x_{n}$ is a \emph{straight
  cycle} if $x_{n} = x_{0}$, $x_{i} \orth x_{i +1}$ for $i = 0,\ldots
,n-1$, $x_{i} \not\comp x_{j}$ whenever $i, j \in \set{0,\ldots ,n-1}$
and $i \neq j$. As usual, the integer $n$ is the length of the cycle.
Observe that a straight cycle is simple, i.e., a part from the
endpoints of the cycle, it does not visit twice the same vertex . The
height of a straight cycle $C = x_{0}x_{1}\ldots x_{n}$ is the integer
\begin{align*}
  \heightp(C) & = \sum_{i = 0,\ldots ,n-1} \heightp(x_{i})\,,
\end{align*}
where $\heightp(x) = \height(x) + 1$ is the augmented height of an
event. By assigning to each element of the cycle a non zero weight, we
can ensure that if $C'$ is another straight cycle visiting a proper
subset of the vertexes visited by $C$, then $\heightp(C') <
\heightp(C)$. This will apply for example when $C'$ is obtained from
$C$ as a shortcut through a chord.

\begin{proposition}
  The graph $\G(\E)$ does not contain a straight cycle of length
  strictly greater than $3$.
\end{proposition}
\begin{proof}
  Let $\CYf$ be the collection of straight cycles in $\G(\E)$ whose
  length is at least $4$. We shall show that if $C \in \CYf$, then
  there exists $C' \in \CYf$ such that $\heightp(C') < \heightp(C)$.
  If $\CYf \neq \emptyset$, then we construct an infinite
  descending chain of positive integers.

  Let $C$ be the straight cycle $x_{0}\orth x_{1}\orth x_{2}\ldots
  x_{n-1} \orth x_{n} = x_{0}$ where $n \geq 4$. Let us suppose that
  this cycle has a chord. Such a chord cut the cycle into two cycles
  of length $m_{0} + 1$ and $m_{1} + 1$, with $m_{0} + m_{1} = n$.  By
  Lemma \ref{lemma:avoided}, we cannot have $m_{0}= m_{1} = 2$, and
  therefore $m_{i} \geq  3$ for some $i \in \set{0,1}$.
  That is, the chord divides the cycle into two straight cycles, one
  of which still has length at least $4$. Moreover its height is
  strictly less than the height of $C$, since it contains a smaller
  number of vertices.

  Otherwise $C$ has no chord and $x_{0}\north x_{2}$. This means that
  either there exists $x'_{0} < x_{0}$ such that $x'_{0}\nconc x_{2}$,
  or there exists $x'_{2} < x_{2}$ such that $x_{0}\nconc x'_{2}$.  By
  symmetry, we can assume the first case holds. As in the proof of
  Lemma \ref{lemma:avoided} $\set{x'_{0},x_{1},x_{2},x_{3}}$ form an
  antichain, and $x'_{0}x_{1}x_{2}x_{3}$ is a path.  Let $C'$ be the
  set $\set{x'_{0}x_{1},\ldots x_{n-1}x_{0}'}$. If $C'$ is an
  antichain, then $C'$ is a straight cycle such that $\heightp(C') <
  \heightp(C)$.  Otherwise the set $\set{j\in \set{4,\ldots
      ,n-1}\mid x_{j} \geq x'_{0}}$ is not empty;  let $i$ be the
  minimum in this set. Observe that $x_{i-1}\orth x_{i}$ and
  $x'_{0} \leq x_{i}$ but $x'_{0} \not\leq x_{i-1}$ implies
  $x_{i-1}\orth x'_{0}$.  Thus $\tilde{C} =
  x'_{0}x_{1}x_{2}x_{3}\ldots x_{i-1} x'_{0}$ is a straight cycle of
  lenght at least $4$ such that $\heightp(\tilde{C}) < \heightp(C)$.
\end{proof}

\begin{corollary}
  \label{cor:antichain}
  Any subgraph of $\G(\E)$ induced by an antichain can be colored with
  $3$ colors.
\end{corollary}
\begin{proof}
  Since the only cycles have length at most $3$, such an induced graph
  is chordal and its clique number is $3$. It is well known that the
  chromatic number of chordal graphs equals their clique number
  \cite{fulkerson}.  
\end{proof}

In the rest of this section we exploit the previous observations to
construct upper bounds for the index of $\E$. We remark that these
upper bounds might appear either too abstract or trivial.  On the
other hand, they well illustrate the obstacles that might arise when
trying to build complex event structures of index  greater than $4$.

A \emph{stratifying function} for $\E$ is a function $h : E \rTo \N$
such that, for each $n \geq 0$, the set $\set{x \in E\mid h(x) = n}$
is an antichain. The height function is a stratifying function. Also
$\varsigma(x) = \card{\set{y \in E\mid y < x}}$
is a stratifying function.  With respect to a stratifying function $h$
the $h$-skewness of $\E$ is defined by
\begin{align*}
  \skewn[h](\E) & = \max\set{|h(x) - h(y)|\mid x\orth y}\,.
  \intertext{More generally, the skewness of $\E$ is defined by}
  \skewn(\E) & = \min\set{%
    \skewn[h](\E)\mid h \text{ is a stratifying function }}\,.
\end{align*}

\begin{proposition}
  \label{prop:skew}
  If $\skewn(\E) < n$ then $\chr(\G(\E)) \leq 3n$.
\end{proposition}
\begin{proof}
  Let $h$ be a stratifying function such that $|h(x) - h(y)| < n$
  whenever $x \orth y$.  For each $k \geq 0$, let $\lambda_{k} :
  \set{x\in E \mid h(x) = k} \rTo \set{a,b,c}$ be a coloring of the
  graph induced by $\set{x\in E \mid h(x) = k}$. Define $\lambda :
  E \rTo \set{a,b,c}\times \set{0,\ldots ,n -1}$ as follows:
  \begin{align*}
    \lambda(x) & = (\lambda_{h(x)}(x), h(x)\mymod n)\,.
  \end{align*}
  Let us suppose that $x \orth y$ and $h(x) \geq h(y)$, so that $0
  \leq h(x) - h(y) < n$. If $h(x) = h(y)$, then by construction
  $\lambda_{h(x)}(x) = \lambda_{h(y)}(x)\neq \lambda_{h(y)}(y)$.
  Otherwise $h(x) > h(y)$ and $0 \leq h(x) - h(y) < n$ implies $h(x)
  \mymod n \neq h(y) \mymod n$. In both cases we obtain $\lambda(x)
  \neq \lambda(y)$.  
\end{proof}

An immediate consequence of Proposition \ref{prop:skew} is the
following upper bound for the index of $\E$:
\begin{align*}
   \chr(\E) & \leq 3(\height(\E) + 1)\,.
\end{align*}
To appreciate the upper bound, consider that another approximation to
the index of $\E$ is provided by Dilworth's Theorem
\cite{dilworth}, stating that $\gamma(\G(\E)) \leq \width(\E)$.
To compare the two bounds, consider that there exist event structures
of degree $3$ whose width is an exponential function of the height.

Finally, we observe that in order to obtain a constant upper bound on
some class of event structures, we can simply define the class
\begin{align*}
  \mathcal{K}_{n}^{k} & = \set{\E \in \mathcal{K}_{n} \mid \max_{x,y
      \in E}|\height(x)
    - \height(y)| < k }\,,
\end{align*}
so that Proposition \ref{prop:skew} ensures that
$\chr(\mathcal{K}_{3}^{k}) \leq 3k$. For $n > 3$, it can still be
shown that $\chr(\mathcal{K}_{n}^{k}) < \infty$, even if the upper
bounds available are not so tight as for $n = 3$.  Let us observe
that the condition
\begin{align*}
  \max_{x,y \in E}|\height(x) - \height(y)| & < k
\end{align*}
appears to be quite natural for an event structure $\E$. An
interpretation of this condition in terms of concurrent processes
appears in the work \cite{madhusudan} whose main purpose is to study
the logical theories of infinite regular event structures.


\section{An Optimal Nice Labelling for Trees and Forests}
\label{sec:thetheorem}

We prove in this section the main contributions of this paper,
Theorems \ref{prop:labelling} and \ref{theo:indexoftrees}. Assuming
$\langle E,\leq\rangle$ is a tree or a forest, we shall define a
labelling with $3$ colors and prove it is a nice labelling.  Since
clearly we can construct a tree which needs at least three colors,
such a labelling is optimal.

Before defining the labelling, we shall develop some observations
about events having the same lower covers. These observations hold
under the assumption that the degree of $\E$ is at most $3$.
\begin{definition}
  We say that two distinct events are \emph{brothers} if they have the
  same set of lower covers.
\end{definition}
Clearly if $x,y$ are brothers, then $z < x$ if and only if $z < y$.
More important, if $x,y$ are brothers, then the relation $x \orth y$
holds.  As a matter of fact, if $x' < x$ then $x' < y$, hence $x'
\wconc y$.  Similarly, if $y' < y$ then $y' \wconc x$. It follows that
a set of events having the same lower covers form a clique in
$\G(\E)$, hence it has at most the degree of an event structure, $3$
in the present case. To introduce the next Lemmas, if $x \in E$, define
\begin{align*}
  \Fam_{x} & = \set{z \in E \mid z \orth x \tand
    y \leq z, \text{ for some brother } y \text{ of } x }\,, \\
  \Soc_{x} & = \set{z \in E \mid z \orth x \tand y \not\leq z, 
    \text{ for every brother } y \text{ of } x
  }\,.
\end{align*}
That is, we are splitting the neighborhood of $x$ into its
\emph{$\Fam$amily}, those events that are descendant of a brother of
$x$, and its \emph{$\Soc$ociety}, those events that are related to $x$
but have no immediate connection with $x$. Intuitively, events in the
family of $x$ are at least as old as $x$, and this will limit our
interest in the family. We shall instead engage in studying properties
of the societies.

\begin{lemma}
  \label{lemma:brothersthree}
  If $x$ has two brothers,
  then $\Soc_{x} = \emptyset$.
\end{lemma}
\begin{proof}
  Let $y,z$ be the two brothers of $x$.  Let us suppose that $w \in
  \Soc_{x}$. If $w \orth y$, then $w \comp z$ by Lemma
  \ref{lemma:avoided}. Since $z \not\leq w$, then $w < z$.  However
  this implies $w < x$, contradicting $w \orth x$.  Hence $w \north y$
  and we can find $w' \leq w$, $y' \leq y$ such that $w' \mconfl
  y'$. It cannot be the case that $y' < y$, otherwise $y' < x$ and the
  pair $(w',y')$, properly covered by the pair $(w,x)$, cannot be a
  minimal conflict. Thus $w' < w$, and $y'$ equals to $y$.  We claim
  that $w' \in \Soc_{x}$. As a matter of fact, $w'$ cannot be above
  any of $x,y,z$, otherwise $w$ would have the same property. From $w
  \orth x$ and $w' < w$, we deduce that $w' \orth x$ or $w' \leq
  x$. If $w' \leq x$, then $w' < x$, so that $w' < y$, contradicting
  $w' \mconfl y$: therefore $w' \orth x$ and $w' \in
  \Soc_{x}$. Observe now that $\set{w',x,y},\set{x,y,z}$ are two
  $3$-cliques sharing the same face $\set{x,y}$. As before, $w' \comp
  z$, leading to a contradiction.
\end{proof}
\begin{lemma}
  \label{lemma:brothers}
  If $y$ is the only brother of $x$, then
  $\Soc_{x},\Soc_{y}$ are comparable w.r.t. subset inclusion and the
  least of them, $\Soc_{x} \cap \Soc_{y}$, is linearly ordered by the
  causality relation.
\end{lemma}
\begin{proof}
  We observe first that if $z \in \Soc_{x}$ and $w \in \Soc_{y}$ then
  $z \comp w$. An immediate consequence of this observation is that
  $\Soc_{x} \cap \Soc_{y}$ is linearly ordered.
  \proofbreath
  Let us suppose that there exists $z \in
  \Soc_{x}$ and $w \in \Soc_{y}$ such that $z \not\comp w$. Note that
  $\set{z,x,y,w}$ is an antichain: $y \not\leq z$, and $z < y$ implies
  $z < x$, which is not the case due to $z \orth x$. Thus $z \not\comp
  y$ and, similarly, $w \not\comp x$.  
  \proofbreath
  Since $z \orth x
  \orth y \orth w$ and there cannot be a length $4$ straight cycle, we
  deduce $z \north w$.  Let $z' \leq z$ and $w' \leq w$ be such that
  $z' \mconfl w'$.  We claim first that $z' \orth x$. Otherwise, $z'
  \leq x$ and $z' < x$, since $z' = x$ implies $x \leq z$. The
  relation $z' < x$ in turn implies $z' < y$, which contradicts $z'
  \mconfl w'$.  Also it cannot be the case that $y \leq z'$, since
  otherwise $y \leq z$.  Thus, we have argued that $z' \in
  \Soc_{x}$. Similarly $w' \in \Soc_{y}$. As before
  $\set{z',x,y,w'}$ is an antichain, hence $z',x,y,w'$ also form a
  length $4$ straight cycle, a contradiction.

  We observe next that $w \leq z \in \Soc_{x}$ and $w \not\leq
  x$ implies $w \in \Soc_{x}$. From $w \leq z \orth x$ deduce $w
  \orth x$ or $w \leq x$.  Since $w \not\leq x$, then $w \orth
  x$. Also, if $y \leq w$ then $y \leq z$, which is not the case.

  Our final observation is that $\Soc_{x} \supseteq \Soc_{y}$ whenever
  $\Soc_{x} \setminus \Soc_{y} \neq \emptyset$.  Let $z \in \Soc_{x}
  \setminus \Soc_{y}$, pick any $w \in \Soc_{y}$ and recall that $z,w$
  are comparable.  We cannot have $z \leq w$: considering that $z
  \not\leq x$, we deduce that $z \not\leq y$ as well; then $z \leq w
  \in \Soc_{y}$ and $z \not\leq y$ imply $z \in \Soc_{y}$, a
  contradiction.  Hence $w < z \in \Soc_{x}$ and $w \not\leq x$ imply
  $w \in \Soc_{x}$, by our previous observation.
\end{proof}
The previous Lemmas have the following interpretation. If $x,y$ are
two brothers, say that \emph{$x$ is more experienced than $y$} if
$\Soc_{x} \supseteq \Soc_{y}$.  Then the Lemmas state that we can
always pick one of the brother who's more experienced than the
other. Remark that the property is trivial if $x,y,z$ are pairwise
brothers, since in this case $\Soc_{w} = \emptyset$, $w \in
\set{x,y,z}$. The property becomes interesting whenever $y$ is the
only brother of $x$, for which we formally introduce this relation.
\begin{definition}
  We say that $\couple{x,y} \subseteq E$ is a \emph{proper pair of
    brothers} if $y$ is the only brother of $x$.
\end{definition}
The next Lemma is an easy consequence of the previous Lemmas. While
its significance might appear obscure right now, the Lemma will prove
to be the key observation when later defining a nice labelling.
\begin{lemma}
  \label{lemma:Oinclusion}
  Let $x,y,z,w \in E$ be four events such that:
  \begin{enumerate}
  \item $\couple{x,y}$ $\couple{z,w}$ are two proper pairs of
    brothers,
  \item $w \not\leq x$,
  \item $z \in \Soc_{x} \cap \Soc_{y}$.
  \end{enumerate}
  Then $z$ is strictly more experienced than $w$, that is
  $\Soc_{z} \supset \Soc_{w}$.
\end{lemma}
\begin{proof}
  If $\Soc_{z} \not\supset \Soc_{w}$, then $\Soc_{z}
  \subseteq \Soc_{w}$ by Lemma \ref{lemma:brothers}.
  If $w \leq y$, then either $w = y$ or $w < y$. We cannot have $w =
  y$, since we are assuming that $w,y$ are distinct. We cannot either
  have $w < y$, since otherwise $w < x$, contradicting $w \not\leq x$.
  Hence we have $w \not\leq y$ and $x,y \in \Soc_{z} \subseteq
  \Soc_{w}$.  It follows that $\set{x,y,z,w}$ is a size $4$
  clique, a contradiction.
\end{proof}


\breath

We come now to introduce trees, that are the particular subsets of $E$
for which we shall define a nice labelling with $3$ letters. The
presence in trees of many brothers possibly is the intuitive reason
for a nice labelling with $3$ letters to exist.
\begin{definition}
  A subset $T \subseteq E$ is a \emph{tree} if and only if
  \begin{enumerate}
  \item each $x \in T$ has exactly one lower cover $\pi(x) \in E$,
  \item $T$ is convex: $x,z \in T$ and $x < y < z$ implies $y \in T$,
  \item if $x,y$ are minimal in $T$, then $\pi(x) = \pi(y)$.
  \end{enumerate}
\end{definition}
The \emph{height} of $x$ in $T$, noted $\height[T](x)$, is the
cardinality of the set $\set{y \in T \mid y < x}$.  Observe that two
events $x,y$ of a tree are brothers if and only if $\pi(x) = \pi(y)$.

In this context, for a linear ordering we shall
mean a transitive \emph{irreflexive} relation $\lord$ which, moreover,
is total: $x \lord y$ or $x = y$ or $y \lord x$.  A linear ordering
$\lord$ on a tree $T$ is said to be compatible with the height if it
satisfies
\begin{align}
  \label{cond:height}
  \tag{HEIGHT}
  \height[T](x) < \height[T](y) & \timplies x \lord y\,.
\end{align}
It is a standard result that such a linear ordering always
exists. Once fixed such a linear ordering, we shall think of it as
imposing a precise age on events of $T$; that is, the relation $x
\lord y$ shall be read as asserting that $x$ is older than
$y$. Observe that the condition \eqref{cond:height} implies that an
ancestor $x$ of $y$ is older than $y$.

With the idea of defining a labelling of $T$ greedily by means of a
fixed linear ordering $\lord$, let us define
\begin{align*}
  \O_{x} & = \set{y \in T \mid  y \orth x \tand y \lord x }\,,
  \;\;\;\;x \in T.
\end{align*} 
That is, $\O_{x}$ is the neighborhood of $x$ within $T$, restricted to
older events.  We represent $\O_{x}$ as the disjoint union of $\C_{x}$
and $\L_{x}$ where
\begin{align*}
  \C_{x} & = \set{y \in \O_{x}\mid \pi(x) < y}\,,
  \\
  \L_{x} & = \O_{x} \setminus \C_{x}\,.
\end{align*}
With respect to these sets $\C_{x},\L_{x}$, $x \in T$, we develop a
series of observations.
\begin{lemma}
  \label{lemma:atmostthree}
  If $y \in \C_{x}$ then $y$ is an older brother of $x$. Consequently
  there can be at most two elements in $\C_{x}$.
\end{lemma}
\begin{proof}
  If $y \in \C_{x}$, then $y \lord x$ and $\height[T](y) \leq
  \height[T](x)$. Since $\pi(x) < y$ then $\height[T](\pi(x)) <
  \height[T](y)$ and $\height[T](x) = \height[T](\pi(x)) + 1 \leq
  \height[T](y)$. We deduce therefore that $\height[T](x) =
  \height[T](y)$, showing that $\pi(x)$ is a lower cover of
  $y$, so that $y$ is a brother of $x$.
\end{proof}
\begin{lemma}
  \label{lemma:LandO}
  $\L_{x}$ is a lower set of $\Soc_{x}$. That is, $\L_{x} \subseteq
  \Soc_{x}$ and $z' \leq z \in \L_{x}$ implies $z' \in \L_{x}$,
  provided that $z' \in \Soc_{x}$.
\end{lemma}
\begin{proof}
  If $z \in \L_{x}$ then $z \orth x$ and $\pi(x) \not\leq z$, hence
  $\pi(x) \not\leq z$. If $y$ is a brother of $x$, then relation $y
  \leq z$ implies $\pi(x) = \pi(y) \leq y \leq z$ and contradicts $z
  \orth \pi(x)$. Hence $y \not\leq z$ and $z \in \Soc_{x}$. Let us
  suppose that $z' < z$ and $z' \orth x$.  Then $\height[T](z') <
  \height[T](z)$, $z' \lord z \lord x$, and $z' \lord x$. Since $z'
  \orth x \geq \pi(x)$ then either $z' \orth \pi(x)$, or $\pi(x) \leq
  z'$. However, the latter property implies $\pi(x) \leq z$, which is
  not the case. Therefore $z' \orth \pi(x)$ and $z' \in \L_{x}$.
\end{proof}

\begin{lemma}
  \label{lemma:Lempty}
  If both $\C_{x}$ and $\L_{x}$ are not empty, then $\C_{x}$ is a
  singleton $\set{y}$ and $\couple{x, y}$ is a proper pair of
  brothers.
\end{lemma}
\begin{proof}
  By the previous Lemma, $\L_{x} \subseteq \Soc_{x}$. Hence, if
  $\L_{x}$ is not empty, then $\Soc_{x}$ is not empty as well, so that
  by Lemma \ref{lemma:brothersthree} $x$ can have at most one
  brother. Since $\C_{x}$ is not empty, and every element in $\C_{x}$
  is a brother of $x$, then $\C_{x}$ has a unique element $y$, and
  $\set{x,y}$ form a proper pair of brothers.
\end{proof}

Let us remark that $x,y \in T$ are a \emph{proper pair of brothers} if
they are brothers and $\set{z\mid \pi(z) = \pi(x)\,} = \set{x,y}$.
The previous observations suggest to look for a linear order $\lord$
that enforces a strictly more experience brother to be an eldest
brother.
\begin{definition}
  We say that a linear order $\lord$ on $T$ is compatible with proper
  pair of brothers if it satisfies \eqref{cond:height} and moreover
  \begin{align}
    \tag{BROTHERS}
    \label{cond:brothers}
    \Soc_{x} \supset \Soc_{y}
    & \timplies x \lord y\,,
  \end{align}
  for each proper pair of brothers $x,y$.
\end{definition}
Again, it is not difficult to see that such a linear order always
exists. In the following we shall assume that $\lord$ satisfies both
\eqref{cond:height} and \eqref{cond:brothers}.

\vskip 12pt

We are ready to define a partial labelling $\lambda$ of the event
structure $\E$.  The function $\lambda$ will have $T$ as its domain.
Let us fix a three elements totally ordered alphabet $\Sigma =
\set{a_{0},a_{1},a_{2}}$.  The labelling $\lambda : T \rTo \Sigma$ is
formally defined by the clauses (1)-(4) to follow.

Before introducing the formal definition, let us introduce some ideas
-- as well as some terminology -- that might help understanding the
definition of $\lambda$ and the proof of Theorem \ref{prop:labelling}.
W.r.t. the linear order $\lord$, we shall say that $x \in T$ is an
\emph{eldest brother} if $\C_{x} = \emptyset$; otherwise, we shall say
that $x$ is a \emph{younger brother}. The clauses (1)-(2) may be
understood as stating that an eldest brother $x$ inherits the property
$\lambda(\pi(x))$ of his father $\pi(x)$. This stipulation will never
create conflicts. The main concern when defining the labelling is to
understand how younger brothers can enrich themselves -- that is, get
a property from the set $\Sigma$ -- without entering in conflict with
members of their neighborhood.  Clause (3) observes that if at the
date of his birth a younger brother is related to his older brothers
only, then these brothers can be at most two and it won't be a problem
getting an unused letter from the alphabet $\Sigma$. Clause (4) is the
subtlest. If at his birth a younger brother $x$ has some relation
outside his family, then he has just one brother $y$ who, by condition
\eqref{cond:brothers}, is more experienced than $x$. In particular, we
shall see that the society of $x$ has a main ancestor $z_{0}$ and is a
main lineage of $z_{0}$, meaning that all of its members are eldest
descendants of $z_{0}$. Thus, assuming that such a lineage has
inherited the same color from its ancestor, we shall see that the
colors used in the neighborhood of $x$ are just $2$; an unused color
from $\Sigma$ is therefore still available.

\begin{definition}
  The labelling $\lambda : T \rTo \Sigma$ is
  formally defined by induction on $\lord$ by following  clauses:
  \begin{enumerate}
  \item If $x \in T$ is an eldest brother and $\height[T](x) = 0$,
    then we let $\lambda(x) = a_{0}$.
  \item If $x \in T$ is an eldest brother and $\height[T](x) \geq 1$,
    let $\pi(x)$ be its unique lower cover. Since $\pi(x) \in T$ and
    $\pi(x) \lord x$, $\lambda(\pi(x))$ is defined and we let
    $\lambda(x) = \lambda(\pi(x))$.
  \item If $x$ is a younger brother and $\L_{x} = \emptyset$, then, by
    Lemma \ref{lemma:atmostthree}, we let $\lambda(x)$ be the least
    symbol not in $\lambda(\C_{x})$.
  \item If $x$ is a younger brother and $\L_{x} \neq \emptyset$ then:
    \begin{itemize}
    \item by Lemma \ref{lemma:Lempty} $\C_{x} = \set{y}$ is a singleton
      and $\couple{x,y}$ is a proper pair of brothers,
    \item by Lemma \ref{lemma:LandO} $\L_{x}$ is a lower set of
      $\Soc_{x}$. By the condition \eqref{cond:brothers}, $\Soc_{x}
      \subseteq \Soc_{y}$, so that $\Soc_{x}$ is a linear order.  Let
      therefore $z_{0}$ be the common least element of $\L_{x}$ and
      $\Soc_{x}$.
    \end{itemize} 
    We let $\lambda(x)$ be the unique symbol not in
    $\lambda(\set{y,z_{0}})$.
  \end{enumerate}
\end{definition}

\begin{theorem}
  \label{prop:labelling}
  For each $x,y \in T$, if $x \orth y$ then $\lambda(x) \neq
  \lambda(y)$.
\end{theorem}
\begin{proof}
  It suffices to prove that $\lambda(y) \neq \lambda(x)$ if $y \in
  \O_{x}$. The statement is proved by induction on $\lord$. Let us
  suppose the statement is true for all $z \lord x$.
  
  (i) If $\height[T](x) = 0$ then $x$ is minimal in $T$, so that
  $\O_{x} = \C_{x}$. If moreover $x$ is an eldest brother then $\O_{x} =
  \C_{x} = \emptyset$, so that the statement holds trivially.

  (ii) If $x$ is an eldest brother and $\height[T](x) \geq 1$, then
  its unique lower cover $\pi(x)$ belongs to $T$.  Observe that $\O_{x}
  = \L_{x} = \set{y\in T\mid y \lord x \tand y \orth \pi(x)}$, so that
  if $y \in \O_{x}$, then $y \orth \pi(x)$.  Since $y \lord x$ and
  $\pi(x) \lord x$, and either $y \in \O(\pi(x))$ or $\pi(x) \in
  \O_{y}$, it follows that $\lambda(x) = \lambda(\pi(x)) \neq
  \lambda(y)$ from the inductive hypothesis.

  (iii) If $x$ is a younger brother and $\L_{x} = \emptyset$, then $\O_{x} =
  \C_{x}$ and, by construction, $\lambda(y) \neq \lambda(x)$ whenever
  $y \in \O_{x}$.

  (iv) If $x$ is a younger brother and $\L_{x} \neq \emptyset$, then
  let $\C_{x} = \set{y}$ and let $z_{0}$ be the common least element of
  $\L_{x}$ and $\Soc^{x,y}_{x}$.  Since by construction $\lambda(x)
  \neq \lambda(y)$, to prove that the statement holds for $x$, it is
  enough to pick $z \in \L_{x}$ and argue that $\lambda(z) \neq
  \lambda(x)$.  We claim that each element $z \in \L_{x} \setminus
  \set{z_{0}}$ is an eldest brother.  If the claim holds, then
  $\lambda(z) = \lambda(\pi(z))$, so that $\lambda(z) =
  \lambda(z_{0})$ is inductively deduced.
  \proofbreath
  Suppose therefore that there exists $z \in \L_{x} \setminus
  \set{z_{0}}$ which is not an eldest brother and let $w \in
  \C_{z}$. Recall first from Lemma \ref{lemma:Lempty} that
  $\couple{x,y}$ form a proper pair of brothers. Similarly,
  $\couple{w,z}$ form a proper pair of brothers. Otherwise, if $z,w,u$
  are pairwise distinct brothers, then either $w \leq x$ or $u \leq x$
  by Lemma \ref{lemma:brothersthree}. In both cases, however, we
  obtain $z_{0} < x$ -- since $z_{0} < z,u,w$ -- which contradicts
  $z_{0} \orth x$. Clearly, $x,y,z,w$ are pairwise distinct as well.
  \proofbreath
  Since $y \lord x$, condition \eqref{cond:brothers} implies $\Soc_{x}
  \subseteq \Soc_{y}$, and hence $z \in \Soc_{x} \cap \Soc_{y}$. If $w
  \in \C_{z}$, then we cannot have $w \leq x$, since again we would
  deduce $z_{0} \leq x$.  Thus we deduce that $w \not\leq x$ and we
  can apply Lemma \ref{lemma:Oinclusion} to deduce $\Soc_{z} \supset
  \Soc_{w}$.  On the other hand, $w \lord z$ and condition
  \eqref{cond:brothers} imply $\Soc_{z} \subseteq \Soc_{w}$.
  Thus, we have reached a contradiction by assuming $\C_{z} \neq
  \emptyset$. It follows that $z$ is an eldest brother. 
\end{proof}

The obvious corollary of Proposition \ref{prop:labelling} is that if
$\E$ is already a sort of tree, then it has a nice labelling with $3$
letters. We state this fact as the following Theorem, after we have
made precise the meaning of the phrase ``$\E$ is a sort of tree.''
\begin{definition}
  Let us say that $\E$ is a \emph{forest} if every element has at most
  one lower cover. Let $\mathcal{F}_{3}$ be the class of event
  structures of degree $3$ that are forests.
\end{definition}
 
\begin{theorem}
  \label{theo:indexoftrees}
  The index of the class $\mathcal{F}_{3}$ is $3$.
\end{theorem}
As a matter of fact, let $\E$ be a forest, and consider the event
structure $\E_{\bot}$ obtained from $\E$ by adding a new bottom
element $\bot$. Remark that the graph $\G(\E_{\bot})$ is the same
graph as $\G(\E)$ apart from the fact that an isolated vertex $\bot$
has been added.  The set of events $E$ is a tree within $\E_{\bot}$,
hence the graph induced by $E$ in $\G(\E_{\bot})$ can be colored with
three colors. But this graph is exactly $\G(\E)$.

To end this section, we mention that Theorem \ref{theo:indexoftrees},
stating the equality between the index and the degree for forests of
degree $3$, does not generalize to forests in higher degrees
\cite{rozoy}.


\section{More Upper Bounds}
\label{sec:w2}

We present in this section some concluding remarks that are meant to
suggest some promising path toward a general solution of the nice
labelling problem for event structures of dgree $3$.

The results presented in the previous sections point out a remarkable
property of event structures of degree $3$: many types of subsets of
events induce a subgraph of $\G(\E)$ that can be colored with $3$
colors.  These include:
\begin{enumerate}
\item \emph{antichains}, by Corollary \ref{cor:antichain},
\item \emph{trees} by Theorem
  \ref{prop:labelling},
\item \emph{history-aware configurations}, since if $X
  \in \Hi$, then $\width(X) \leq 3$, so that such a subset can be
  labeled with $3$ letters by Dilworth's Theorem,
\item the \emph{stars} of events.
\end{enumerate}
The star of an event $x \in E$ is the subgraph of $\G(\E)$
induced by the subset $\set{x} \cup \set{y \in E\mid y \orth x}$. To
see that a star can also be labeled with $3$ letters, let
\begin{align*}
  N_{x} & = \set{y \in E \mid y \orth x } \intertext{be the
    neighborhood of $x$ in $\G(\E)$, and consider the structure}
  \E_{x} & = \langle N_{x},\leq_{|N_{x}},\CNF_{|N_{x}} \rangle\,,
\end{align*}
where $\leq_{|N_{x}}$ is the restriction of $\leq$ to $N_{x}$ and
$\CNF_{|N_{x}} = \set{X \cap N_{x} \mid X \in \CNF}$.
\begin{lemma}
  $\E_{x}$ is a coherent event structure with the property that 
  $\G(\E_{x})$ is the subgraph of $\G(\E)$ induced by $N_{x}$.
 Consequently $\cl(\E_{x}) < \cl(\E)$.
\end{lemma}
\begin{proof}
  We leave the reader to verify that $\E_{x}$ is an event structure
  whose concurrency relation $\conc_{x}$ is  the
  restriction of $\conc$ to the set $N_{x}$. Consequently
  $\CNF_{|N_{x}}$ is the set of cliques for $\conc_{x}$ and $\E_{x}$
  is coherent.  Let $y \orth_{x} z$ be the orthogonality relation of
  $\E_{x}$, let us verify that, for $y,z \in N_{x}$, $y \orth_{x}
  z$ if and only if $y \orth z$.

  If $y \orth z$ then $y,z$ are not comparable. If $y' \in N_{x}$ and
  $y' < y$, then either $y' \leq z$ or $y' \conc z$, that is $y'
  \conc_{x} z$. By symmetry, $z' < z$ with $z' \in N_{x}$ implies 
  $z' \leq y$ or $y \conc_{x} z'$, thus $y \orth_{x} z$.
  \proofbreath
  Let us suppose in the other direction that $y \orth_{x} z$.  Then
  $y,z$ are not comparable. If $y' < y$ and $y' \in N_{x}$,
  then $y' \conc_{x} z$ or $y' \leq z$, which implies $y'\wconc z$.
  If $y' < y$ but $y' \not\in N_{x}$, then $y' < x$.  From $z \orth x$
  and $y' < x$ it follows that $y' \wconc x$. Similarly, if $z' < z$
  then $z' \wconc x$ and therefore we can deduce  $y \orth z$.

  Finally, observe that, by adding the event $x$, a size $n$ clique in
  $\G(\E_{x})$ gives rise to a size $n +1$ clique in $\G(\E)$.
\end{proof}
We finalize our discussion by observing that if $\cl(\E) = 3$, then
$\cl(\E_{x}) \leq 2$ so that $\E_{x}$ has a labelling with $2$
letters, by \cite{rozoy}. It follows that star of $x$, the $\set{x}
\cup N_{x}$, can be labeled with $3$ letters.

\breath

We might ask whether this property can be exploited to construct
nice labellings.  A tentative answer comes from a standard technique
in graph theory \cite{zykov}. Consider a partition ${\cal P} =
\set{[z]\mid z \in E}$ of the set of events such that each equivalence
class $[z]$ has a labelling with $3$ letters. Define the quotient
graph $\G({\cal P},\E)$ as follows: its vertexes are the equivalence
classes of ${\cal P}$ and $[x] \orth \relax [y]$ if and only if there
exists $x' \in [x]$, $y' \in [y]$ such that $x' \orth y'$.
\begin{proposition}
  \label{prop:quotient}
  If the graph $\G({\cal P},\E)$ is $n$-coloriable, then $\E$ has a
  labelling with $3n$ colors. 
\end{proposition}
\begin{proof}
  For each equivalence class $[x]$ choose a coloring $\lambda_{[x]}$
  of $[x]$ with an alphabet with $3$ letters. Let $\lambda_{0}$ a
  coloring of the graph $\G({\cal P},\E)$ and define $\lambda(x) =
  (\lambda_{[x]}(x),\lambda_{0}([x]))$. Then $\lambda$ is a coloring
  of $\E$: if $x \orth y$ and $[x] = [y]$, then $\lambda_{[x]}(x) =
  \lambda_{[y]}(x) \neq \lambda_{[y]}(y)$ and otherwise, if $[x] \neq
  [y]$, then $[x] \orth\relax [y]$ so that $\lambda_{0}([x]) \neq
  \lambda_{0}([y])$.  
\end{proof}
The reader should remark that the technique suggested by Proposition
\ref{prop:quotient} has already been used within Proposition
\ref{prop:skew}.

We conclude our discussion by exemplifying how to use the Labelling
Theorem for trees \ref{prop:labelling} in connection with Proposition
\ref{prop:quotient} to construct a finite upper bound for the index of
a particular class of event structures. This class shall be called
simple due to the additional simplifying properties of its structures.

\begin{figure}
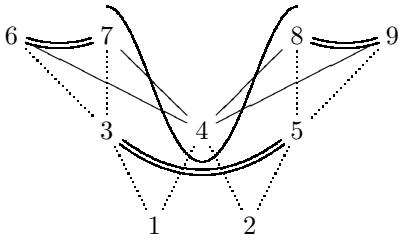

  \centering
  $$
  \xygraph{
    []*+{1}="1"([r]*+{2})
    [l(0.5)u]*+{3}="3"([r]*+{4}="4"[r]*+{5}="5")
    "3"[l(1)u]*+{6}="6"([r]*+{7}="7")
    "5"[l(0)u]*+{8}="8"([r]*+{9}="9")
    "1"(:@{.}"3",:@{.}"4")
    "2"(:@{.}"4",:@{.}"5")
    "3"(:@{.}"6",:@{.}"7")
    "5"(:@{.}"8",:@{.}"9")
    "6":@/_0.4em/@{=}"7"
    "8":@/_0.4em/@{=}"9"
    "3":@/_1.6em/@{=}"5"
    "6":@{-}"4"
    "7":@{-}"4"
    "8":@{-}"4"
    "9":@{-}"4"
    "7"[u(0.3)]="START"
    "8"[u(0.3)]="END"
    "START"[r(0.2)]="A"
    "4"[l(0.4)]="B"
    "4"[d(0.5)]="C"
    "4"[r(0.4)]="D"
    "END"[l(0.2)]="E"
    "START"
    -@`{"A","B","C","D","E"}
    "END"
  }
  $$  
  \caption{The event structure $\mathcal{S}$}
\end{figure}
Consider the event structure depicted in Figure 1, named ${\cal
  S}$. In this picture we have used 
dotted lines for the edges of the Hasse diagram of $\langle E,\leq
\rangle$, simple lines for maximal concurrent pairs, and double lines
for minimal conflicts.  Concurrent pairs $x \conc y$ that are not
maximal, i.e.  for which there exists $x',y'$ such that $x' \orth y'$
and either $x< x'$ or $y < y'$, are not drawn. We leave the reader to
verify that a nice labelling of ${\cal S}$ needs at least $4$
letters. 
On the other hand, it is not difficult to find  a nice
labelling with $4$ letters. To obtain it, take apart events
with at most $1$ lower cover from the others, as suggested in the
picture.
Use then the results of the previous section to label with three
letters the elements with at most one lower cover, and label the only
element with two lower covers with a forth letter.

\breath

A formalization of this intuitive method leads to the following
Definition and Proposition.
\begin{definition}
  We say that an event structure is \emph{simple} if 
  \begin{enumerate}
  \item it is graded, i.e.  $\height(x) = \height(y) - 1$ whenever $x
    \lcover y$,
  \item every size $3$ clique of $\G(\E)$ contains a minimal conflict.
  \end{enumerate}
\end{definition}
The event structure ${\cal S}$ is simple and proves that even simple
event structures cannot be labeled with just $3$ letters.
\begin{proposition}
  Every simple event structure $\E$ of degree $3$ has a nice labelling
  with $12$ letters.
\end{proposition}
\begin{proof}
  Recall that $\fat(x)$ is the number of lower covers of $x$ and let
  $E_{n} = \set{ x \in E \mid \fat(x) = n}$. Observe that a simple
  $\E$ is such that $E_{3} = \emptyset$: if $x \in E_{3}$, then its
  three lower covers form a clique of concurrent events. Also, by
  considering the lifted event structure $\E_{\bot}$, introduced at
  the end of section \ref{sec:thetheorem}, we can assume that
  $\card(E_{0})=1$, i.e.  $\E$ has just one minimal element which
  necessarily is isolated in the graph $\G(\E)$.

  Let $\lord$ be a linear ordering of $E$ compatible with the height.
  W.r.t. this linear ordering we shall use a notation analogous to the
  one used in the previous section.  We let 
  \begin{align*}
    \O_{x} & = \set{y \in E\mid y
      \lord x \tand y \orth x}\,,& 
    \B_{x} &  = \set{y \in E\mid  y
      \orth x \tand y' \lcover y \timplies y' \lcover x}\,.
  \end{align*}
  \begin{claim}
    The subgraph of $\G(\E)$ induced by $E_{2}$ can be colored with
    $3$-colors.
  \end{claim}
  We remark first that if $x \in E_{2}$ then $\O_{x} \subseteq
  \B_{x}$. Let $y \in \O_{x}$ and let $x_{1},x_{2}$ be the two lower
  covers of $x$.  From $x_{i} < x \orth y$ it follows $x_{i} < y$ or
  $x_{i} \conc y$.  If $x_{i} \conc y $ for $i =1,2$, then
  $y,x_{1},x_{2}$ is a clique of concurrent events. Therefore, at
  least one lower cover of $x$ is below $y$, let us say $x_{1} < y$.
  It follows that $\height(y) \geq \height(x)$, and since $y \lord x$
  implies $\height(y) \leq \height(x)$, then $x,y$ have the same
  height. We deduce that $x_{1} \lcover y$. If $y$ has a second lower
  cover $y'$ which is distinct from $x_{1}$, then $y' = x_{2}$,
  otherwise $y',x_{1},x_{2}$ is a clique of concurrent events.
  \proofbreath Next, we remark that if $y,z \in \B_{x}$ and $x \in
  E_{2}$ then $y \orth z$: if $y' < y$ then $y' \leq x$ so that $x
  \orth z$ implies $y' \wconc z$, and symmetrically. It follows that
  for $x \in E_{2}$, $\B_{x}$ may have at most $2$ elements. 
  \proofbreath A fortiori, $\O_{x}$ has at most $2$ elements which
  always form a clique. 
  The restriction of $\lord$ to $E_{2}$ is therefore a $2$-elimination
  ordering by which we can color $E_{2}$ with $3$ colors.
  \eproofofclaim

  For $x \in E_{1}$ let $\rho(x) = \max \set{z \in E \mid z \leq x, z
    \not\in E_{1}}$ and $[x] = \set{y \in E_{1} \mid \rho(y) =
    \rho(x)}$.
     Let ${\cal P}$ be the partition
      $\set{E_{0}} \cup \set{[x] \mid x \in E_{1}} \cup
      \set{E_{2}}$.
  Since each $[x]$, $x \in E_{1}$, is a tree, the partition ${\cal P}$
  is such that each equivalence class induces a $3$-colorable subgraph
  of $\G(\E)$.
  
  \begin{claim}
    The graph $\G({\cal P},\E)$ is $4$-colorable.
  \end{claim}
  \takebreath
  Since $E_{0}$ is isolated in $\G({\cal P},\E)$, is it enough to
  prove that the subgraph of $\G({\cal P},\E)$ induced by the trees
  $\set{[x] \mid x \in E_{1} }$ is $3$-colorable.  
  We define fist a linear ordering $\lord$ on the set of trees by
  stating that $[y] \lord [x]$ if and only if $\rho(y) \lord \rho(x)$.
  \proofbreath
  As usual, let $\O_{[x]} = \set{[y] \orth [x] \mid [y] \lord [x]}$, we
  claim that $\O_{[x]}$ may contain at most two trees.  To this goal,
  we shall define a function $f : \O_{[x]} \rTo \B_{\rho(x)}$ and
  prove it is injective.
  If $[y] \orth\relax [x]$ and $[y] \lord
  [x]$ then we can pick $y' \in[y]$ and $x' \in [x]$ such that $y'
  \orth x'$. Notice that $y' \orth \rho(x)$: from $\rho(x) \leq x'
  \orth y'$, we deduce $\rho(x) \orth y'$ or $\rho(x) \leq y'$. The
  latter, however, implies $\rho(x) \leq \rho(y)$, by the definition
  of $\rho$, and this relation contradicts $\rho(y) \lord
  \rho(x)$. Thus we let
  \begin{align*}
    f([y]) & = \min \set{z \mid \rho(y) \leq z \leq y' \tand z
      \not\leq \rho(x)}\,.
  \end{align*}
  This definition implies that $f([y]) \orth \rho(x)$: as a matter of
  fact, $f([y]) \leq y' \orth \rho(x)$ and $f([y]) \not\leq \rho(x)$
  implies $f([y]) \orth \rho(x)$.  Moreover every lower cover of
  $f([y])$ is a lower cover of $\rho(x)$: this statement clearly holds if
  $f([y]) \neq \rho(y)$, and if $f([y]) = \rho(y)$ then it holds since
  $\rho(y) \lord \rho(x)$ implies $f([y]) \in \O_{\rho(x)} \subseteq
  \B_{\rho(x)}$, as in the proof of the previous Claim.
  \proofbreath Thus the set $f(\O_{[x]})$ has cardinality at most $2$
  and, to prove that $\O_{[x]}$ has at most $2$ elements, we prove
  that $f$ is injective.  The intuitive reason is that $f$ is a choice
  function, i.e. $f([y]) \in \set{\rho(y)} \cup [y]$.  Let us suppose
  that $f([y]) = f([z])$.  If $f([y]) = \rho(y)$, then $f([z]) =
  \rho(z)$ as well and $[y] = [z]$.  Otherwise $f([y]) = f([z])$
  implies $\rho(y) = \rho(f([y])) = \rho(f([z])) = \rho(z)$ and $[y] =
  [z]$.  \eproofofclaim
  
  Thus, by applying Proposition \ref{prop:quotient}, we deduce that
  $\G(\E)$ has a labelling with $12$ letters.  
\end{proof}



\bibliographystyle{elsart-num}
\bibliography{biblio}

\end{document}